\documentclass{kapproc} 
\usepackage{epsf}

\normallatexbib

\begin{document}
\articletitle[Berryonic Matter in the Cuprates]{Berryonic Matter in the 
Cuprates}
\author{R.S. Markiewicz}
\affil{Physics Department and Barnett Institute, Northeastern U., Boston, 
MA 02115}


\begin{abstract}
     A novel form of Jahn-Teller (JT) effect in the cuprates can be 
reinterpreted as a conventional JT effect on a lattice with a larger unit cell. 
There is a triplet of instabilities, parametrized by a pseudospin, consisting 
of a form of the low-temperature tetragonal phase, a 
charge density wave phase, and a flux phase (orbital antiferromagnet).  On a
single 4-Cu plaquette, the problem is of $E\otimes (b_1+b_2)$ form.  For a 
special choice of parameters, the model supports a dynamic JT effect, but is 
classically chaotic.  The connection of this phase with Berryonic matter is 
discussed.
\end{abstract}

For stripe phases in the cuprates, an important question is why the charged 
stripes are metallic: why do they have a preferred 
doping $x_0<1$ hole per Cu?  One attractive possibility is that optimal
doping corresponds to fixing the Fermi level at the Van Hove singularity (VHS).
Then an electronic instability would gap a large density of states, making the
holes nearly incompressible and minimizing the electronic free energy 
(`Stability from Instability.')  It has been proposed that the dominant 
electronic instability could be to a charge-density wave (CDW) like 
state\cite{RM3,Cast,Surv}.  Here I explore the phonon anomalies that might
accompany this CDW.

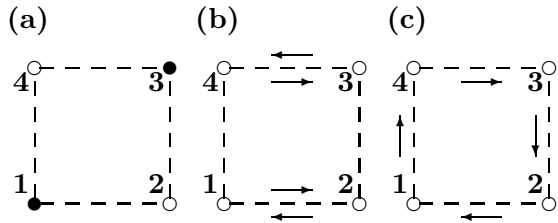
\begin{figure}\begin{center}
\setlength{\unitlength}{0.9mm}
\begin{picture}(80,40)
\footnotesize
\put(2,10){\circle*{2}}
\put(22,10){\circle{2}}
\put(2,30){\circle{2}}
\put(22,30){\circle*{2}}
\multiput(2.8,10)(3.8,0){5}{\line(2,0){2}}
\multiput(2.8,30)(3.8,0){5}{\line(2,0){2}}
\multiput(2,10.8)(0,3.8){5}{\line(0,2){2}}
\multiput(22,10.8)(0,3.8){5}{\line(0,2){2}}
\put(30,10){\circle{2}}
\put(30,30){\circle{2}}
\put(50,30){\circle{2}}
\put(50,10){\circle{2}}
\multiput(30.8,10)(3.8,0){5}{\line(2,0){2}}
\multiput(30.8,30)(3.8,0){5}{\line(2,0){2}}
\multiput(30,10.8)(0,3.8){5}{\line(0,2){2}}
\multiput(50,10.8)(0,3.8){5}{\line(0,2){2}}
\put(37,28){\vector(1,0){6}}
\put(43,32){\vector(-1,0){6}}
\put(37,12.){\vector(1,0){6}}
\put(43,8.){\vector(-1,0){6}}
\put(58,10){\circle{2}}
\put(78,10){\circle{2}}
\put(58,30){\circle{2}}
\put(78,30){\circle{2}}
\multiput(58.8,10)(3.8,0){5}{\line(2,0){2}}
\multiput(58.8,30)(3.8,0){5}{\line(2,0){2}}
\multiput(58,10.8)(0,3.8){5}{\line(0,2){2}}
\multiput(78,10.8)(0,3.8){5}{\line(0,2){2}}
\put(65,28){\vector(1,0){6}}
\put(71,8.){\vector(-1,0){6}}
\put(76,23){\vector(0,-1){6}}
\put(56,17){\vector(0,1){6}}
\normalsize
\put(1,37){\makebox(0,0){\bf (a)}}
\put(29,37){\makebox(0,0){\bf (b)}}
\put(57,37){\makebox(0,0){\bf (c)}}
\put(20,28){\makebox(0,0){\bf 3}}
\put(20,13){\makebox(0,0){\bf 2}}
\put(0,13){\makebox(0,0){\bf 1}}
\put(0,28){\makebox(0,0){\bf 4}}
\put(48,28){\makebox(0,0){\bf 3}}
\put(48,13){\makebox(0,0){\bf 2}}
\put(28,13){\makebox(0,0){\bf 1}}
\put(28,28){\makebox(0,0){\bf 4}}
\put(76,28){\makebox(0,0){\bf 3}}
\put(76,13){\makebox(0,0){\bf 2}}
\put(56,13){\makebox(0,0){\bf 1}}
\put(56,28){\makebox(0,0){\bf 4}}
\end{picture}
\caption{Pseudospin triplet: CDW (a), LTT (b), OAF (c).}
\label{fig:1}
\end{center}\end{figure}

There are three related CDW-like distortions which couple to the VHS, all of 
which have been proposed to play a role in the cuprates.  These distortions, 
which form a pseudospin triplet\cite{MarV}, are illustrated in Fig.~\ref{fig:1}.
The CDW couples to the oxygen breathing mode, which was early proposed as a
driving force for high $T_c$\cite{Web} (a dynamic instability of the related
half-breathing mode is associated with the stripes\cite{McQ1}).  The LTT 
distortion couples to the strain associated with the low-temperature tetragonal
(LTT) phase, which is implicated in pinning the stripes\cite{Tran}.  Finally,
the orbital antiferromagnet (OAF) is closely related to the flux phase, a 
candidate in many theories of cuprate physics\cite{Affl,Laugh}.

Before the role of stripes was understood, a model was proposed of the low
temperature orthorhombic (LTO) phase as a dynamic LTT phase, related to a novel
Van Hove -- Jahn-Teller (VHJT) effect\cite{RM8c,Surv}, where the two VHS's
constitute the electronic degeneracy. A much deeper understanding of both the
phonon anomalies and the VHJT effect can be gained by considering the cuprates
as a Berryonic solid\cite{Berr}, with the CuO$_2$ planes composed of a square 
array of Cu$_4$O$_8$ molecules.  Each 2$\times$2 plaquette is Jahn-Teller
(JT) active, so the VHJT effect on the original lattice is equivalent to a
conventional JT effect on the plaquette lattice.  Taking one hole per Cu, the
electronic states on a plaquette can be symmetrized to yield states of $A_{1g}$,
$B_{2g}$, and $E_u$ symmetry.  For the lattice as a whole, {\it all states near 
the VHS's are built up exclusively of $E_{u}$ states}\cite{RM8c}.  These are
the states closest to the Fermi level, and doping the isolated plaquette leads
to a conventional $E\otimes (b_1+b_2)$ JT problem\cite{JT2}.  The 
B$_1$ mode couples to the LTT, the B$_2$ to the CDW, and the dynamic JT phase 
corresponds to the flux phase.

Here, I study one particular question: in JT terms, under what circumstances
could the flux (dynamic JT) phase appear in the plaquette?  The relevant parts 
of the Hamiltonian are the phonon and JT terms:
\begin{equation}
H_{ph}={1\over 2M}\bigl(P_1^2+P_2^2+M^2\omega_1^2Q_1^2+M^2\omega_2^2Q_2^2\bigr),
\label{eq:5b}
\end{equation}
\begin{equation}
H_{JT}=V_1Q_1T_x+V_2Q_2T_y,
\label{eq:5}
\end{equation}
with bare phonon frequencies $\omega_i$, electron-phonon coupling $V_i$, and 
electronic pseudospins $T_i$ representing the $E_u$ states.  The JT energy is 
$E_{JT}^{(i)}=V_i^2/(2\omega_i^2)$.  
When $E_{JT}^{(1)}\ne E_{JT}^{(2)}$, the ground state is a static JT distortion,
corresponding to the state with larger $E_{JT}$.  More interesting is the case
when $E_{JT}^{(1)}=E_{JT}^{(2)}$.  If in addition $\omega_1=\omega_2$, the
problem reduces exactly to the well known $E\otimes e$ problem, with anomalous 
Berry phase\cite{CAM} signifying the dynamic JT ground state.  

\begin{figure}
\begin{center}
\leavevmode
   \epsfxsize=0.7\textwidth\epsfbox{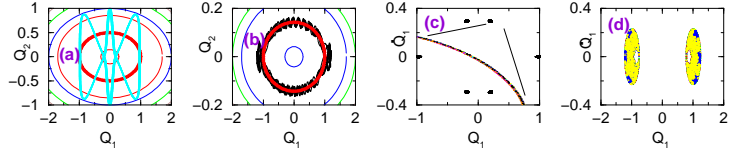}
\vskip0.5cm 
\caption{(a,b): Time series, $Q_2(t)$ vs $Q_1(t)$ for $(\omega_2/\omega_1,\beta 
)$: (a) = (2,1), (b) = (7,0.05394).  Ellipses are equipotential contours, with 
the beaded contour representing the potential minimum.  (c,d): Corresponding
Poincare maps. In (c), one attractor is shown on expanded scale.}
\label{fig:43}
\end{center}
\end{figure}

However, in the cuprates the phonon frequencies $\omega_1$ (LTT) and $\omega_2$ 
(breathing mode) are 
very different.  Nevertheless, since $V_2>V_1$, it is possible that $E_{JT}^{(1)
}=E_{JT}^{(2)}$ may hold approximately in the cuprates.  Curiously, this special
case is not well studied.  It appears to have a Berry phase (and hence the 
dynamic JT effect), but is also chaotic. Since the vibronic potential has an 
elliptic minimum, circulating orbits should be possible, even though the angular
momentum $j_z$ is not constant.  When the classical Hamiltonian (particle in a 
non-linear potential well) is numerically integrated, the generic solution is
found to be chaotic: the `particle' gets partially trapped near one of the 
extrema of the ellipse, and after a delay can be reflected or transmitted at 
random.  Some typical trajectories are illustrated in Fig.~\ref{fig:43}a,b, 
while the corresponding Poincare maps (plots of $Q_1$ vs $\dot Q_1$ when $Q_2=0
$) are in Figure~\ref{fig:43}c,d.  The data sets are parametrized by $(\omega_2 
/\omega_1,\beta )$, with $\beta$ being a scaled initial velocity.  For some 
special initial conditions, nearly periodic solutions can be found, but even for
these the Poincare sections are weakly chaotic.  Note that the chaos is present 
even though the frequencies are rationally related.  

\begin{figure}\begin{center}
\leavevmode
   \epsfxsize=0.4\textwidth\epsfbox{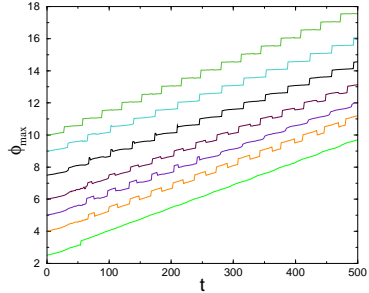}
\vskip0.5cm 
\caption{Solution to one-dimensional version of quantum JT 
problem, showing evolution of $\phi_{max}$ (the value
of $\phi$ at which the probability density is largest) vs $t$, for several 
values of frequency anisotropy: from bottom to top, $\alpha$ = 0.03, 0.04, 0.06,
0.1, 0.3, 0.6, 1.  Different curves are shifted by assuming different initial 
positions of the wave function.}
\label{fig:34}
\end{center}\end{figure}

The quantum problem has also been analyzed, but so far only in a one-dimensional
limit, in which the motion is confined to the bottom of the trough and only 
$\phi$ varies.  By rescaling the potentials, the potential can be made to have
circular symmetry, with anisotropic effective masses.  Schroedinger's equation 
can be integrated numerically, and it is found that the quantum system
shows a 
`memory' of the classical chaos.  Figure~\ref{fig:34} shows the time evolution 
of the most probable $\phi$ value as a function of time.  The wave function 
remains trapped most of the time in one of the effective potential wells (near 
the points $Q_2=0$), then quickly hops to the next one in a relatively
short time.  However, the tunneling is coherent, so there is a net circulation.
In Fig.~\ref{fig:34}, results for a number of different values of anisotropy 
$\alpha\equiv (\omega_2^2-\omega_1^2)/(\omega_2^2+\omega_1^2)$ are shown.  
Disregarding some transients, the general trend is
that the steps get washed out as the isotropic ($\alpha\rightarrow 0$)
limit is approached, but $\phi$ always contains a monotonic, nearly 
$\alpha$-independent component. Hence, in this case also, a flux-phase-like 
state exists.

\end{document}